\documentclass[9pt,twocolumn,twoside]{osajnl}

\journal{ol} 

\setboolean{shortarticle}{true}

\newcommand{\bra}[1]{\left<#1\right|}      
\newcommand{\ket}[1]{\left|#1\right>}      

\newcommand{\df}[2]{\frac{\partial #1}{\partial #2}} 
\newcommand{\cra}[1]{\hat{a}^{\dag}_{#1}}  
\newcommand{\ana}[1]{\hat{a}_{#1}}         

\title{Probing the topology of the two-photon bands via time-dependent quantum walks}
\author[1,*]{Andrei~A.~Stepanenko}
\author[1]{Maxim~A.~Gorlach}
\affil{Department of Physics and Engineering, ITMO University, Saint Petersburg 197101, Russia}

\affil[*]{Corresponding author: andrey.stepanenko@metalab.ifmo.ru}

\begin{abstract}
Topological protection of quantum correlations opens new horizons and opportunities in quantum technologies. A variety of topological effects has recently been observed in qubit networks. However, the experimental identification of the topological phase still remains challenging, especially in the entangled many-body case. Here, we propose an approach to independently probe single- and two-photon topological invariants from the time evolution of the two-photon state in a one-dimensional array of qubits. Extending the bulk-boundary correspondence to the two-photon scenario, we show that an appropriate choice of the initial state enables the retrieval of the topological invariant for the different types of the two-photon states in the interacting Su-Schrieffer-Heeger model. Our analysis of the Zak phase  reveals additional facets of topological protection in the case of collapse of bound photon pairs.
\end{abstract}

\setboolean{displaycopyright}{true}

\begin{document}

\maketitle

The concept of topological order that relies on the behavior of eigenstates in the Brillouin zone has become attractive due to the accompanying disorder-robust topologically protected states~\cite{Hasan2010, Ozawa2019} realized experimentally at various platforms~\cite{Rechtsman2013,Hafezi2013,Ma2015}. Further application of topological platforms requires a method to identify the topological phase in an experiment. Historically, first approaches defined the topological phase based on the characteristic edge state \cite{Mourik2012, Hart2014} or on the measurement of a spin texture \cite{Hsieh2009}. However, rapid growth of experimental realizations of topological systems required new techniques for the direct extraction of the topological invariant.

The topological invariant is a quantity that is conserved during adiabatic change of the system parameters, which does not cause the bandgap to close. Previously, the Chern number (two-dimensional topological invariant) was extracted from the center of mass motion of an ultracold Fermi~\cite{Dauphin2013} or Bose~\cite{Aidelsburger2014} gas.
At the same time, 
an experimental probe of a one-dimensional topological invariant (Zak phase) was performed using a combination of Bloch oscillations and Ramsey interferometry for cold atoms in an optical lattice \cite{Atala2013}. Another method that relies on the measurement of mean chiral displacement temporal evolution was proposed for Floquet systems~\cite{Cardano2017} and later applied to waveguide arrays~\cite{Longhi2018} and superlattices~\cite{Longhi2019} followed by the theoretical analysis for chiral models~\cite{Maffei2018}, and photon quantum walks in qubit arrays~\cite{Ramasesh2017, Flurin2017, Mei2018, Cai2019}.

Superconducting qubits~\cite{Kjaergaard2020} is a promising platform for the study of quantum phenomena which has already been employed for the demonstration of topological effects for quantum states~\cite{Roushan2016,Roushan2017,Cai2019,Kim2021,Besedin2020}. While physics of the single-particle quantum topological states is relatively well-understood, the focus of research shifts to the study of quantum many-body topological states and associated correlations~\cite{BlancoRedondo2018, Mittal2018, Wang2019} taking into account interaction effects \cite{Smirnova2020,Liberto2016,Bello2017,Gorlach2017,Marques2017,Zurita2019} which requires the technique to extract topological invariants in quantum many-body regime.

In this Letter, we investigate temporal dynamics of the different types of the two-photon states including bound photon pairs and scattering states. Exploring the evolution of the mean chiral displacement with time, we outline the approach to extract the topological invariant (Zak phase) for the two-photon quantum states and test this algorithm for the special cases such as collapse of bound photon pair.

\par
\begin{figure}[t]
\centering
\includegraphics[width=\linewidth]{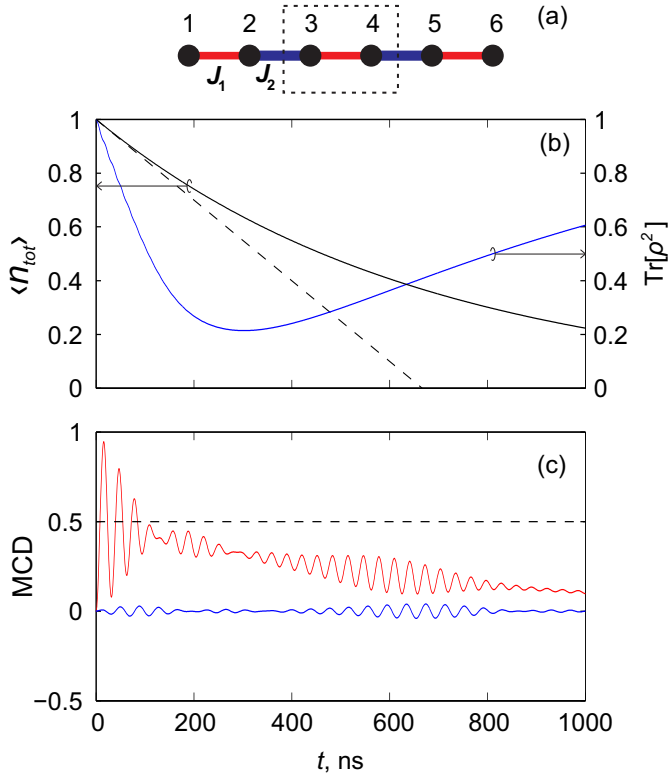}
\caption{ (a) The scheme of the Su-Schrieffer-Heeger model with a unit cell shown in black dashed rectangle. (b) Evolution of the  single-photon state. Black curve shows the dynamics of the total number of photons. Black dashed line illustrates the characteristic relaxation time $\sim 1/\gamma_\downarrow$. Blue line shows  Tr$[\rho^2]$ indicating the purity of the quantum state. (c) Mean chiral displacement for topological (red) and trivial (blue) cases calculated for the system with parameters $\gamma_\downarrow = 1.5$ MHz, $\gamma_\phi = 10$ MHz, $J_2/J_1 = 10$, $f = 3.8$ GHz  and the length of the array $2N=22$.
}
\label{fig:1}
\end{figure}

 As a specific example, we analyze a one-dimensional array of  qubits with the alternating coupling amplitudes that corresponds to the Su-Schrieffer-Heeger (SSH) model~\cite{Su1979} recently realized for qubit arrays~\cite{Cai2019,Kim2021,Besedin2020}. The considered system is depicted in Fig.\ref{fig:1}(a) and described by the Bose-Hubbard Hamiltonian
\begin{eqnarray}
\label{eq:Hamiltonian}
  \hat{H} &=& f\,\sum_{m}\hat{n}_m+ U\sum_{m}\hat{n}_m(\hat{n}_m-1)
  -  J_1 \sum_{m}(\cra{2m} \ana{2m-1}+\textrm{H.c.})\nonumber\\ && 
  - J_2 \sum_{m}(\cra{2m} \ana{2m+1}+ \textrm{H.c.})
  \:,
\end{eqnarray}
where $\hat{n}_i$ is the photon number operator and $\ana{i}$ is the annihilation operator. We set $h = 1$ for simplicity. Note that the nonlinear term $U$ spectrally separates states with two photons in the same qubit and leads to the formation of bound photon pair (doublon) both for attractive ($U<0$) and repulsive ($U>0$) interaction~\cite{Winkler2006}. 

The two-photon states in such Hermitian model have been extensively investigated in Refs.~\cite{Liberto2016,Gorlach2017,Marques2017}. However, the realistic array of qubits has inevitable losses, which give rise to the relaxation along with the decoherence. To describe the temporal dynamics of the two-photon states in such a system, we solve Lindblad equation for the density matrix, taking into account quantum states with zero, one and two elementary excitations:
\begin{eqnarray}
\df{\rho}{t}=-i\left[H,\rho\right]+\gamma_\downarrow \sum_m L[\hat{a}_m](\rho)+\gamma_\phi \sum_m L[\hat{n}_m](\rho)\:,
\end{eqnarray}
\begin{eqnarray}
L[\hat{c}](\rho) &=& \hat{c}\rho\hat{c}^\dagger-\frac{\hat{c}^\dagger\hat{c}\rho+\rho\hat{c}^\dagger\hat{c}}{2}\:.
\end{eqnarray}
Here $\gamma_\downarrow$ and $\gamma_\phi$ are the dissipation and decoherence rates, respectively, and $\hat{c}$ is an arbitrary operator.
Solving this equation, we obtain the density matrix that starts its evolution from the initial state $\rho(0)$ and evolves according to the law $\rho(t)$. Given the density matrix, the expectation values of the observables $\hat{c}$ are calculated as
%
$\langle \hat{c}\rangle = Tr[\rho\hat{c}]$.
Note that tuning of the initial state allows us to select the band for which the topological invariant is extracted.

For clarity, we revisit first the single-photon case featuring a topological state at the weak link edge~\cite{Su1979}. We test the  approach \cite{Cardano2017} for the single-photon case and construct mean chiral displacement (MCD) operator that in a coordinate basis includes only diagonal elements:
\begin{eqnarray}
\hat{M}_{s,s}=(\Gamma \hat{x})_{ss} = (-1)^{j_1-1}(i_1-(N-1)/2), \nonumber\\
s = 2i_1+j_1,\: 0\le i_1\le N-1,\:j_1=1,2\:,
\end{eqnarray}
where $N$ is the total number of the unit cells, index $i_1$ numerates the unit cells and $j_1$ corresponds to the first or second qubit inside the unit cell. Previously, it was shown that the long-time asymptotic of this quantity is related to the Zak phase $\gamma$ as $\langle \hat{M}\rangle = \gamma/(2\pi)$.

This result is well understandable in the fully dimerized limit. Since the dimers are independent, the photon localizes inside the particular dimer. If the dimer is inside the unit cell (topologically trivial case), the particle's position indicated by the unit cell number is an integer and in the opposite case (topologically non-trivial case) it takes half-integer values, e.g. $0.5$.

Next we consider the system shown in Fig.~\ref{fig:1}(a) with the ratio of coupling constants $J_2/J_1 = 10$ chosen such that the weak coupling constant is at the edge and inside the unit cell. Assuming that the excitation is initialized in the middle of the array, we expect to observe a nontrivial topological phase with MCD equal to 0.5. However, the considered system is open and the total photon number decreases with time due to the dissipation [black curve in Fig.\ref{fig:1}(b)] that causes the decrease of the MCD [red line in Fig.\ref{fig:1}(c)] in the topological case. If the ratio of the tunneling constants is inverted ($J_1/J_2 = 10$), the stronger link appears inside the unit cell which corresponds to the topologically trivial case characterized by the oscillations of MCD near zero [blue line in Fig.\ref{fig:1}(c)].

In our simulations, we assume that the system is prepared in the state with a single photon in a single qubit in the middle of the array. While the initial state is pure, it soon becomes mixed due to the decoherence. However, at long evolution times this mechanism competes with the relaxation which increases the weight of the vacuum state bringing the system to the pure vacuum state in the limit $t\rightarrow\infty$. The competition of the two mechanisms explains the characteristic time dependence of Tr$[\rho^2]$ depicted in Fig.\ref{fig:1}(b) by the blue line and featuring a characteristic minimum at the intermediate time scales.

Having discussed the single-photon scenario, we focus on the two-photon case generalizing the approach to measure the topological invariant. In the two-photon sector, the model \eqref{eq:Hamiltonian} supports four types of states which are scattering states; doublons (two photons are co-localized); single-photon edge states (one of the two photons is localized at the edge), and doublon edge states~\cite{Gorlach2017}. Below we demonstrate that the appropriate definition of mean chiral displacement allows one to probe the topological properties of all three families of the two-photon states except for the doublon edge states which arise as a manifestation of topology of the two photon bands.


\subsection{A doublon state}
We start our analysis from doublons which are a distinctive feature of two-photon interacting problem.
Limiting the basis only to the doublon states, we define the doublon MCD operator:
\begin{eqnarray}
\hat{M}_{s,s}=(\Gamma \hat{x})_{ss} = (-1)^{j_1-1}(i_1-(N-1)/2), \nonumber\\
s = ((4N - 2i_1-j_1)(2i_1+j_1 - 1))/2 + 2i_1+j_1, \nonumber\\ 0\le i_1\le N-1,\:j_1=1,2\:.
\label{mcd2}
\end{eqnarray}
%
%
Although it is possible to obtain a rigorous analytical solution for the doublon dispersion in the studied system~\cite{Liberto2016, Gorlach2017}, for the sake of clarity we focus on the limiting case which allows us to clearly reveal the topological features of the model.

We assume that the interaction strength exceeds both of the tunneling constants ($U\gg J_{1,2}$) and therefore doublon can be considered as a single stable quasi-particle. This makes the problem analogous to the single-photon regime. However, the topological phase transition in now controlled by the ratio of the effective coupling constants $J_2^{(\rm{eff})}/J_1^{(\rm{eff})} = J_2^2/J_1^2$. 

\begin{figure}[t]
\centering
\includegraphics[width=\linewidth]{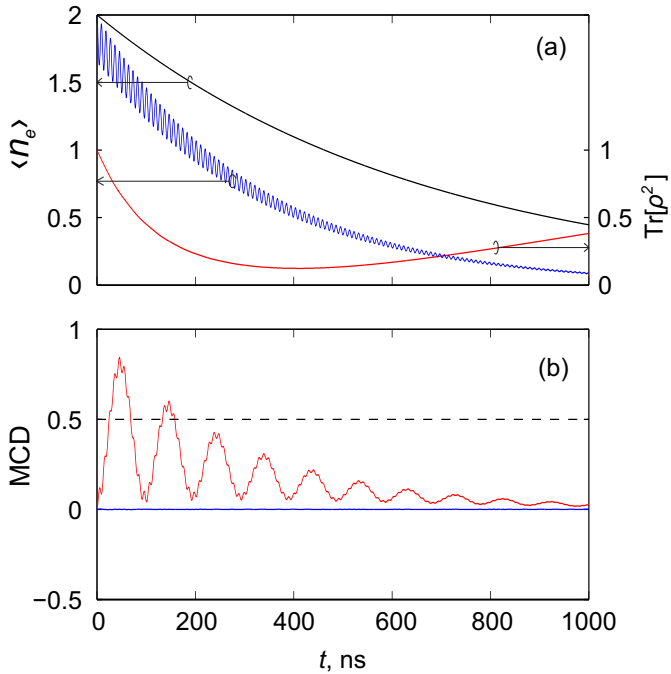}
\caption{Illustration of the doublon dynamics. (a) The black curve shows the dynamics of the total number of photons $\langle\hat{n}_{tot}\rangle$. The blue curve corresponds to the number of doublons multiplied by 2: 2$\langle\hat{n}_{d}\rangle =\langle \hat{a}^\dagger\hat{a}^\dagger\hat{a}\hat{a}\rangle$. The quantity Tr$[\rho^2]$ showing the purity of the quantum state is depicted by the red line. (b) MCD for the topological (red) and trivial (blue) cases. $2N=10$ and $U=-277.5$ MHz The rest of the parameters are the same as in Fig. \ref{fig:1}. 
}
\label{fig:2}
\end{figure}

Tracing the evolution of the photon number [black curve, Fig.~\ref{fig:2}(a)], we observe the situation quite similar to the single-photon case. At the same time, the number of doublons in the system exhibits rapid oscillations combined with a slower trend of gradual decrease that correlates with the decrease of the total photon number [blue curve, Fig.~\ref{fig:2}(a)]. These characteristic oscillations are the consequence of the doublon motion, since doublon propagation implies two consecutive tunneling processes. When one of the photons moves to the adjacent site, the number of doublons defined as $\langle\hat{n}_{d}\rangle =\langle \hat{a}^\dagger\hat{a}^\dagger\hat{a}\hat{a}\rangle/2$ temporarily decreases, while the next tunneling step restores the number of doublons. The behavior of Tr$[\rho^2]$ [red line, Fig.~\ref{fig:2}(a)] illustrates the similar competition of the  dissipation and decoherence as in the single-photon case.

The calculated value of the topological invariant [red curve in Fig.\ref{fig:2} (b)] ocillates and decreases with approximately the same rate as the number of doublons, which highlights the primary role played by the relaxation process in the evolution of topological invariant. In the opposite case corresponding to the trivial choice of the unit cell [blue curve in Fig.\ref{fig:2}(b)] no significant oscillations occur.

The described topological behaviour is accompanied by closing and reopening of the bandgap with the change of the ratio of the tunneling constants~\cite{Liberto2016, Gorlach2017} which illustrates the bulk-boundary correspondence in the two-photon regime~\cite{Gorlach2017,Stepanenko2020a}. 

However, in the case of moderate interactions the dispersion of bound photon pair can intersect with the scattering continuum and hence doublons appear to be stable only in a certain range of wave numbers. Such situation termed here doublon collapse~\cite{Lin2014,Gorlach2017} leaves topological characteristics of doublon undefined since the doublon band does not exist in the entire Brillouin zone.

\begin{figure}[t]
\centering
\includegraphics[width=0.9\linewidth]{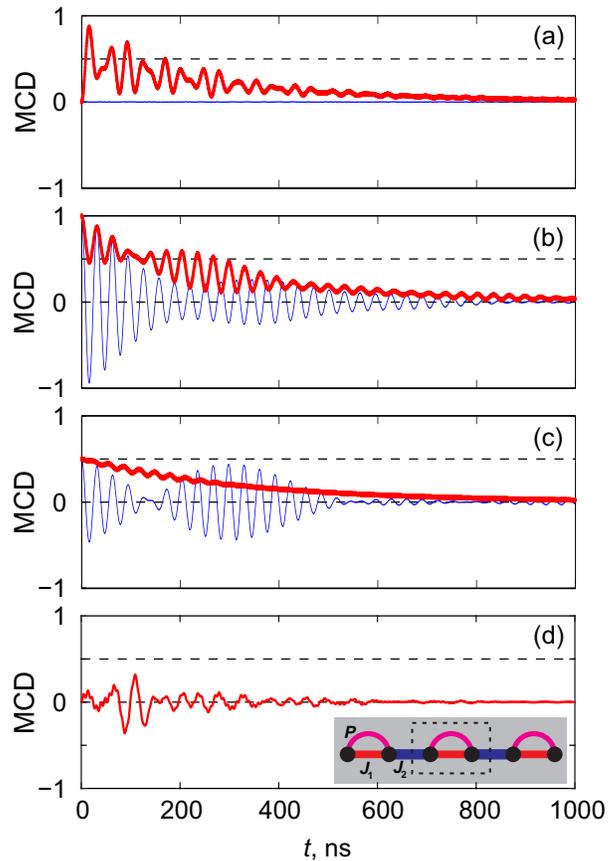}
\caption{Evolution of the different types of two-photon states. (a) MCD for the doublon states in the situation of collapse~-- intersection  of one bulk band with the continuum of scattering states ($U=-37.5$ MHz). Red and blue curves correspond to the topological and trivial scenario, respectively.
(b) MCD for topological (red) and trivial (blue) cases simulated  for the state with edge localization of one of the photons. (c) The same as (b) but for the scattering state. (d) MCD for the doublon state in the extended Bose-Hubbard model with additional direct two-photon tunneling $P=40$~MHz. $2N=10$, $U=-277.5$ MHz, the rest of the parameters are the same as in Fig. \ref{fig:1}.
}
\label{fig:3}
\end{figure}

From the theoretical perspective, it was suggested to extract the topological invariant examining the remaining non-collapsing band~\cite{Gorlach2017, Stepanenko2020a}. Below, we prove that the procedure of topological invariant retrieval based on quantum walks yields a consistent result with this theoretical proposal.

Initializing the two-photon state in the non-collapsing band, we recover the MCD dynamic shown in Fig.\ref{fig:3}(a). The obtained result for MCD is slightly different from that considered before [cf. Fig.\ref{fig:2}(b)] due to the interaction of the different types of the two-photon states. To excite the non-collapsing band, we prepare the system in the superposition of the doublon states $(\ket{2_{N}}\pm \ket{2_{N+1}})/\sqrt{2}$, where $\ket{2_n}$ denotes the two-photon Fock state localized in the $n^{\rm{th}}$ qubit. However, the similar results can be obtained when the initial state is mixed: $\hat{\rho} = (\ket{2_{N}}\bra{2_{N}} + \ket{2_{N+1}}\bra{2_{N+1}})/2$.

\subsection{Two-photon scattering state}

To probe the topological properties of the two-photon scattering states, we fill in the missing elements of the MCD operator in order to take into account all positions of the two photons comprising the pair for the given center-of-mass coordinate:
\begin{eqnarray}
\hat{M}_{s,s}=(\Gamma \hat{x})_{ss} = \frac{1}{2}\left[(-1)^{j_1-1}(i_1-\frac{N-1}{2})\right.\nonumber\\\left.+(-1)^{j_2-1}(i_2-\frac{N-1}{2})\right], 
\end{eqnarray}
where
\begin{eqnarray}
s = \frac{(4N - 2i_1-j_1)(2i_1+j_1 - 1)}{2} + 2i_2+j_2, \nonumber\\ 0\le i_1\le N-1,\:j_1=1,2\:,
\nonumber\\ i_1\le i_2\le N-1,\:j_2=
\left\{
\begin{array}{cc}
    j_1,2\:, & i_1=i_2\:, \\
    1,2\:, & i_1\neq i_2\:
\end{array}
\right.
\end{eqnarray}

Here we again consider the regime of strong anharmonicity so doublon states do not interfere with the scattering continuum. Therefore, we expect that this type of evolution should yield a  single-photon topological invariant. The calculated time evolution of MCD is demonstrated in Fig.~\ref{fig:3}(b,c) for the two distinct cases: one photon is confined to the topological edge state, but the other propagates in the bulk [Fig.~\ref{fig:3}(b)], and two photons propagate in the bulk [Fig.~\ref{fig:3}(c)]. In both cases, MCD decreases with the total photon number due to the dissipation, while the total doublon number exhibits slight oscillations near zero. To study these two situations, we prepare the system in the initial states $\hat{a}_1^\dagger\,\hat{a}_6^\dagger\ket{0}$ and $\hat{a}_4^\dagger\,\hat{a}_6^\dagger\ket{0}$, respectively. Note, that the edge localization can be confirmed measuring the photon number in the edge qubit.

It should be stressed that the system investigated here is attainable in an experiment and can be readily investigated. However, in such setup the topological invariant for the single- and two-photon states is the same and, as a consequence, different types of the two-photon states feature the same asymptotic behavior  of mean chiral displacement.

In order to discriminate single- and two-photon topological invariants, we consider an extended version of the Bose-Hubbard model~\cite{Stepanenko2020a} illustrated in the inset of Fig.\ref{fig:3}(d) and characterized by the additional interaction term in the Hamiltonian \eqref{eq:Hamiltonian} responsible for the direct two-photon tunneling: $\sim P\sum_m(\cra{2m-1}\cra{2m-1}\ana{2m}\ana{2m}+H.c.)$. Since this term is manifested only in the two-photon case, it enables different topological phases in the single- and two-photon cases and allows to highlight the distinctive features of the approach developed here.

Performing the calculation of MCD in this case [Fig.\ref{fig:3}(d)], we show that doublon MCD takes the value different from the single-photon case for the same choice of the unit cell [Fig.~\ref{fig:1}(c)]. The obtained result also contrasts with the calculated values of MCD for the scattering states which remain qualitatively the same as those shown in Fig.~\ref{fig:3}(c).

Even though the experimental implementation of such extended model is not straightforward, our analysis highlights that the single- and two-photon topological invariants can be probed separately via time-dependent quantum walks taking different values under the appropriate conditions.

In summary, we have developed a technique to extract the  two-photon topological invariant in one-dimensional systems  based on the measurement of time evolution of the mean chiral displacement. This technique allows one to independently probe single- and two-photon topological invariants which for some structures can take different values. Our results suggest that the developed method is significantly more sensitive to the dissipation rather than to the decoherence. Furthermore, we prove the applicability of our technique in the situation when one of the bulk doublon bands collapses providing an access to the topology of unstable quasi-particles.









\begin{backmatter}
\bmsection{Funding} 

\bmsection{Acknowledgments} This work was supported by the Russian Foundation for Basic Research (Grant No.~18-29-20037). A.A.S. and M.A.G. acknowledge partial support by the Foundation for the Advancement of Theoretical Physics and Mathematics ``Basis". A.A.S. also acknowledges partial support by the Government of the Russian Federation (Grant No.~SP-3707.2021.5) and partial support by the Gennady Komissarov Foundation.





\bmsection{Disclosures} The authors declare no conflicts of interest.

\bmsection{Data availability} 
The data that supports the findings of this study is available from the corresponding author upon request. 


\end{backmatter}





\bibliography{ol1}

\bibliographyfullrefs{ol1}



\end{document}